%% file: main.tex
\documentclass[sigconf,screen]{acmart}
\usepackage{float}
\usepackage{listings}
\usepackage{xcolor}
\definecolor{mauve}{rgb}{0.58, 0, 0.82}

\definecolor{bg}{rgb}{0.95,0.95,0.95}
\usepackage{listings}
\usepackage{adjustbox}
\graphicspath{{./figures/}}
\DeclareGraphicsExtensions{.png,.pdf}
\definecolor{bg}{rgb}{0.95, 0.95, 0.95}

\usepackage{caption}
\usepackage{mathtools}
\usepackage{mathpartir}
\usepackage{proof}
\usepackage{appendix}
\usepackage{setspace}
\usepackage{comment}
\usepackage{tikz}
\usepackage{lipsum}
\usepackage{xspace}
\usepackage{amsmath}
\usepackage[linesnumbered,ruled,vlined]{algorithm2e}
\usepackage{algpascal}
\usepackage{mdframed}
\usepackage{algc}
\usepackage{algcompatible}
\usepackage{algpseudocode}
\usepackage{listings}
\usepackage{xcolor}
\definecolor{mauve}{rgb}{0.58, 0, 0.82}

\usepackage{bm}
\usepackage{multirow}
\usepackage{enumitem}
\usepackage{stmaryrd}
\usepackage{makecell}
\usepackage{graphicx}
\usepackage{etoolbox}

\makeatletter
\patchcmd{\@algocf@start}{-1.5em}{0pt}{}{}
\makeatother

\definecolor{dkgreen}{rgb}{0,0.6,0}
\definecolor{dkblue}{rgb}{0,0,0.8}

\AtBeginDocument{
  \providecommand\BibTeX{
    {\normalfont B\kern-0.5em{\scshape i\kern-0.25em b}\kern-0.8em\TeX}
  }
}

\newcommand{\accuracyLow}{97.2\%\ }

\newcommand{\compilationtime}{8.6\%\ }

\copyrightyear{2025}
\acmYear{2025}
\setcopyright{cc}
\setcctype{by}
\acmConference[ISSTA Companion '25]{34th ACM SIGSOFT International Symposium on Software Testing and Analysis}{June 25--28, 2025}{Trondheim, Norway}
\acmBooktitle{34th ACM SIGSOFT International Symposium on Software Testing and Analysis (ISSTA Companion '25), June 25--28, 2025, Trondheim, Norway}\acmDOI{10.1145/3713081.3732932}
\acmISBN{979-8-4007-1474-0/2025/06}
\usepackage{balance}

\begin{document}

\title{Towards Source Mapping for Zero-Knowledge Smart Contracts: Design and Preliminary Evaluation}

\author{Pei Xu}
\affiliation{
  \institution{University of Technology Sydney}
  \city{Sydney}
  \country{Australia}
}
\email{pei.xu@student.uts.edu.au}

\author{Yulei Sui}
\affiliation{
  \institution{University of New South Wales}
  \city{Sydney}
  \country{Australia}
}
\email{y.sui@unsw.edu.au}

\author{Mark Staples}
\affiliation{%
\institution{CSIRO’s Data61 and}
  \city{Digital~Finance~CRC}
  \country{Australia}
}
\email{mark@dfcrc.com}

\begin{abstract}
Debugging and auditing zero-knowledge-compatible smart contracts remains a significant challenge due to the lack of source mapping in compilers such as \texttt{zkSolc}. In this work, we present a preliminary source mapping framework that establishes traceability between Solidity source code, LLVM IR, and zkEVM bytecode within the \texttt{zkSolc} compilation pipeline. Our approach addresses the traceability challenges introduced by non-linear transformations and proof-friendly optimizations in zero-knowledge compilation.

To improve the reliability of mappings, we incorporate lightweight consistency checks based on static analysis and structural validation. We evaluate the framework on a dataset of 50 benchmark contracts and 500 real-world zkSync contracts, observing a mapping accuracy of approximately \accuracyLow\ for standard Solidity constructs. Expected limitations arise in complex scenarios such as inline assembly and deep inheritance hierarchies. The measured compilation overhead remains modest, at approximately \compilationtime.

Our initial results suggest that source mapping support in zero-knowledge compilation pipelines is feasible and can benefit debugging, auditing, and development workflows. We hope that this work serves as a foundation for further research and tool development aimed at improving developer experience in zk-Rollup environments.
\end{abstract}



\maketitle

\input{1.introduction}
\input{2.motivating}
\input{3.background}
\input{4.approach}
\input{5.experiment}
\input{6.limitation}
\input{7.related}
\input{8.conclusion}

\bibliography{bibliography}

\end{document}

%% file: 1.introduction.tex
\section{Introduction}
\label{sec:intro}

Blockchain technology has become a cornerstone of decentralized digital asset systems, enabling transparent, secure, and efficient applications~\cite{kushwaha2022ethereum}. Among the solutions addressing blockchain scalability challenges, zk-Rollups have emerged as a promising technology. They allow Ethereum’s Layer 2 ecosystem to process large transaction volumes while preserving strong security guarantees~\cite{xu2022zkrollup}. A key enabler of zk-Rollups is \texttt{zkSolc}, a specialized compiler that extends the Solidity compiler to produce zero-knowledge-compatible bytecode, commonly referred to as \emph{zkEVM bytecode}.

\emph{Source mapping} refers to the process of correlating compiled bytecode instructions with their corresponding high-level source code locations. This mapping is essential for debugging, auditing, and symbolic execution, as it enables developers and auditors to trace execution behavior back to source code statements~\cite{solidity2025sourcemappings}.

While source mapping is a standard feature in Solidity compilers such as \texttt{solc}, it is currently absent in \texttt{zkSolc}~\cite{matterlabs2024sourcemap}. This omission introduces practical challenges for developers working on zero-knowledge-based smart contracts. Specifically, the \texttt{zkSolc} compilation pipeline involves multiple transformation stages, including translation to LLVM IR and subsequent generation of zkEVM bytecode. These stages introduce non-linear transformations, such as instruction reordering, arithmetic simplification, and control flow restructuring, which break the direct correspondence between source code and bytecode offsets that traditional source mapping techniques rely on.

In this work, we propose a preliminary source mapping framework for \texttt{zkSolc}. Our objective is to improve traceability across compilation stages and facilitate debugging and auditing in zk-Rollup environments.

\subsection{Contributions}

This paper makes the following contributions:

\begin{itemize}
    \item We propose a source mapping framework for \texttt{zkSolc} that links Solidity source code, LLVM IR, and zkEVM bytecode, accounting for transformations specific to zero-knowledge compilation.
    \item We introduce a lightweight consistency validation process to assess the structural integrity of the generated mappings.
    \item We report preliminary evaluation results on benchmark and real-world smart contracts, analyzing mapping accuracy and compilation overhead.
\end{itemize}

The remainder of this paper is organized as follows. Section~\ref{sec:motivation} presents a motivating example. Section~\ref{sec:preliminaries} introduces key concepts and defines the problem. Section~\ref{sec:methodology} describes the proposed framework. Section~\ref{sec:experiments} reports evaluation results. Section~\ref{sec:limitations} discusses limitations. Section~\ref{sec:related_work} reviews related work, and Section~\ref{sec:conclusion} concludes the paper.

%% file: 2.motivating.tex
\section{Motivating Example}
\label{sec:motivation}

To illustrate the importance of source mapping in \texttt{zkSolc} and the practical challenges that arise in its absence, we present a debugging scenario involving a privacy-preserving voting contract. This contract uses zero-knowledge proofs (zk-SNARKs) to protect voter anonymity while ensuring vote integrity. A simplified Solidity implementation is shown below:

\begin{figure}[H]
\begin{lstlisting}[
    backgroundcolor=\color{bg}, 
    basicstyle=\small\ttfamily,
    frame=single, 
    breaklines 
]{solidity}
contract ZKVoting {
    mapping(address => bool) public hasVoted;

    function submitVote(bytes memory zkProof) external {
        require(verifyZKProof(zkProof), "Invalid proof");
        require(!hasVoted[msg.sender], "Already voted");

        hasVoted[msg.sender] = true;
    }

    function verifyZKProof(bytes memory zkProof) internal pure returns (bool) {
        // Placeholder for zero-knowledge proof verification logic
        return true; // Simplified for illustration
    }
}
\end{lstlisting}

\subsection{Debugging Challenges Without Source Mapping}

In practice, debugging smart contracts compiled with \texttt{zkSolc} is challenging due to the absence of source mapping. Consider the following scenario: A user submits a vote, but the transaction fails at the \texttt{require} statement in the \texttt{submitVote} function. The error message \texttt{"Invalid proof"} indicates that \texttt{verifyZKProof} returned \texttt{false}. Without source mapping, identifying the root cause is difficult and may stem from:

\begin{enumerate}
    \item A bug in the Solidity implementation of \texttt{verifyZKProof}.
    \item An unintended transformation during the LLVM IR compilation stage.
    \item A miscompilation or instruction reordering in the LLVM IR to zkEVM bytecode translation.
\end{enumerate}

Without a mapping mechanism, developers must manually inspect low-level representations. For example, the corresponding LLVM IR may appear as:

\begin{lstlisting}[ 
    backgroundcolor=\color{bg}, 
    basicstyle=\small\ttfamily,
    frame=single, 
    breaklines 
]
; Check the result of verifyZKProof
%3 = call i1 @verifyZKProof(%bytes* %zkProof)
br i1 %3, label %4, label %5
\end{lstlisting}
\end{figure}

While this snippet shows a conditional branch based on \texttt{verifyZKProof}, it provides no direct connection to the original Solidity source. Similarly, the associated zkEVM bytecode:

\lstdefinelanguage{text}{} 
\begin{figure}[H]
\begin{lstlisting}[language=text, 
    backgroundcolor=\color{bg}, 
    basicstyle=\small\ttfamily, 
    frame=single, 
    breaklines
]
0x03 PUSH1 0x40
0x04 MSTORE
0x05 CALLDATALOAD
\end{lstlisting}
\end{figure}

lacks context or information about its origin in the Solidity logic. This absence of traceability complicates debugging and increases the manual effort required to locate errors.

\subsection{Implications for Security Audits and Optimization}

Beyond debugging, source mapping plays an essential role in security auditing and performance analysis. Auditors inspecting for vulnerabilities such as reentrancy or access control flaws often analyze low-level bytecode to assess contract behavior. Without source mapping, auditors must manually correlate bytecode instructions with Solidity source code, increasing the risk of misinterpretation.

Similarly, developers seeking to optimize gas costs require a clear mapping between expensive bytecode instructions and their high-level Solidity counterparts. For example, a sequence of bytecode instructions such as:

\begin{figure}[H]
\begin{lstlisting}[language=text, 
    backgroundcolor=\color{bg}, 
    basicstyle=\small\ttfamily,
    frame=single, 
    breaklines 
]
0x10 CALLDATASIZE
0x11 ISZERO
0x12 PUSH2 0x0040
0x13 JUMPI
\end{lstlisting}
\end{figure}

could correspond to multiple Solidity statements, making it difficult to attribute gas consumption without source-level context.

\subsection{Challenges in Implementing Source Mapping in \texttt{zkSolc}}

While source mapping is a standard feature in traditional EVM compilers, implementing it in \texttt{zkSolc} presents unique challenges. The \texttt{zkSolc} compilation pipeline introduces multiple transformation stages, including lowering Solidity to LLVM IR, applying proof-friendly optimizations, and generating zkEVM-compatible bytecode. These transformations disrupt the direct correspondence between source code and compiled output.

For instance, a simple bitwise operation in Solidity:

\begin{figure}[H]
\begin{lstlisting}[language=text, basicstyle=\small\ttfamily]
function checkBit(uint256 input) public pure returns (bool) {
    return (input & 1) == 1;
}
\end{lstlisting}
\end{figure}

may be transformed during compilation into arithmetic constraints used in zk-SNARK proof generation:

\begin{figure}[H]
\begin{lstlisting}[language=text, basicstyle=\small\ttfamily]
Constraint 1: tmp1 = input % 2
Constraint 2: tmp2 = tmp1 == 1
\end{lstlisting}
\end{figure}

Here, \texttt{tmp2} is a temporary variable representing an intermediate value generated during IR-level translation of a Solidity expression. These transformations improve proof efficiency but obscure the original source semantics, making conventional source mapping infeasible.

These transformations improve proof efficiency but obscure the original source semantics, making conventional source mapping infeasible.

Additionally, LLVM optimizations such as inlining, constant folding, and dead code elimination further alter or eliminate source-level constructs. Witness generation in zero-knowledge proofs imposes bit-level precision requirements that introduce further rewrites, complicating traceability.

Performance overhead is another consideration. Including detailed source mapping increases the size of compiled artifacts and introduces additional processing steps during compilation, potentially impacting deployment costs and throughput.

Despite these challenges, we believe that source mapping support in \texttt{zkSolc} is an important direction. Improved traceability can assist developers and auditors in understanding contract behavior, identifying errors, and analyzing performance. This work represents an initial attempt to address this gap by proposing a source mapping framework tailored to the \texttt{zkSolc} compilation pipeline.

%% file: 3.background.tex
\section{Preliminaries}
\label{sec:preliminaries}

This section introduces key background concepts relevant to our work, including zk-Rollups, the \texttt{zkSolc} compiler, and the challenges associated with source mapping in zero-knowledge compilation pipelines.

\subsection{Blockchain Scaling and zk-Rollups}

Blockchain scalability remains a central challenge. On Ethereum, limited transaction throughput and high gas fees have driven the development of Layer 2 scaling solutions that execute transactions off-chain while periodically settling state changes on-chain~\cite{LayerTwoSurvey, Rao2024}. Among these, zk-Rollups leverage zero-knowledge proofs to validate the correctness of batched transactions without requiring full re-execution on the base layer.

In a typical zk-Rollup system, transactions are aggregated off-chain, and a succinct zk-SNARK proof is generated to attest to the correctness of the batch. This proof is submitted to an Ethereum smart contract, which verifies it efficiently. Unlike optimistic rollups that rely on fraud proofs and challenge periods, zk-Rollups enforce correctness at the time of state transition, providing immediate finality and strong security guarantees.

\subsection{\texttt{zkSolc}: The Zero-Knowledge Solidity Compiler}

\texttt{zkSolc} is a specialized compiler that extends the standard Solidity compiler to generate bytecode compatible with zero-knowledge proof systems. Specifically, it produces \emph{zkEVM bytecode}, a variant of EVM bytecode optimized for zero-knowledge execution.

Unlike \texttt{solc}, which directly compiles Solidity source code to EVM bytecode, \texttt{zkSolc} introduces an intermediate compilation stage based on LLVM IR~\cite{matterlabs2025toolchain,immutable2022zkevm}. This additional stage enables advanced optimizations tailored to zk-SNARK constraint generation but introduces substantial transformations that complicate traceability between source code and compiled output.

\begin{figure*}[t]
    \centering
    \includegraphics[width=0.75\textwidth]{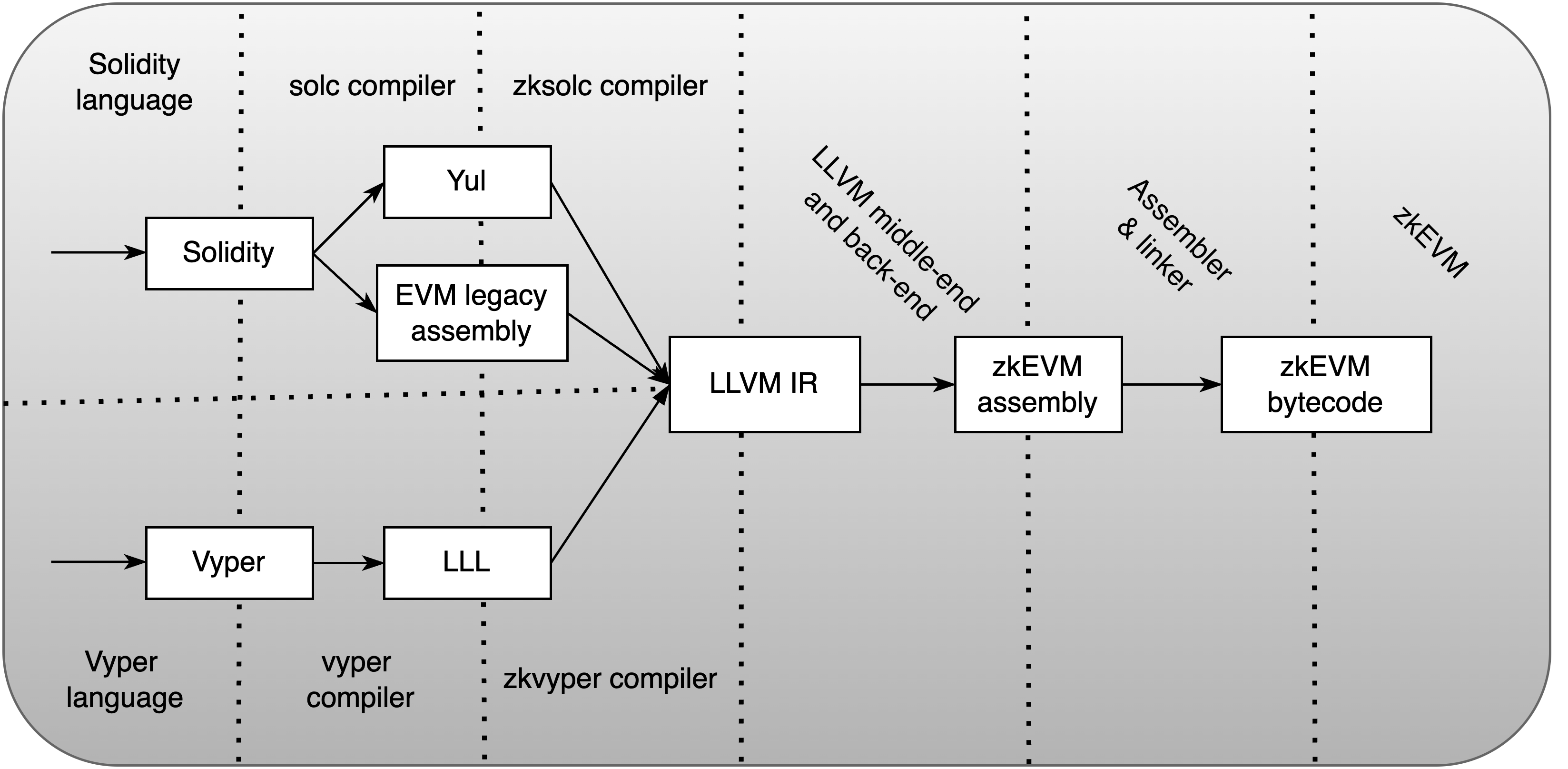}
    \caption{\texttt{zkSolc} compilation pipeline~\cite{zksync_toolchain}. Yul is an intermediate language used in Solidity compilation. Vyper and LLL are alternative smart contract languages supported by EVM toolchains but are outside the scope of this work.}
    \label{fig:compiler}
\end{figure*}

As illustrated in Figure~\ref{fig:compiler}, the \texttt{zkSolc} compilation pipeline consists of the following stages:

\begin{enumerate}
    \item Parsing Solidity source code into an abstract syntax tree (AST).
    \item Lowering the AST to a high-level intermediate representation (Yul).
    \item Translating Yul to LLVM IR in static single assignment (SSA) form.
    \item Applying LLVM optimizations to simplify arithmetic operations, restructure control flow, and minimize circuit constraints.
    \item Generating zkEVM-compatible bytecode suitable for zero-knowledge execution.
\end{enumerate}

These transformations improve proof efficiency but significantly alter instruction structure and semantics compared to the original Solidity source code. As a result, conventional offset-based source mapping techniques used in \texttt{solc} are not directly applicable.

\subsection{Source Mapping in Solidity Compilers}

Source mapping is a standard feature in traditional smart contract compilers, providing a correspondence between compiled bytecode instructions and their source-level origins. In conventional Solidity compilers such as \texttt{solc}, source maps are generated to facilitate debugging, symbolic execution, and security auditing. These maps enable developers to trace execution behavior back to Solidity source lines, supporting tools such as Remix IDE~\cite{remix}, Hardhat Debugger~\cite{hardhat-debug}, and Truffle~\cite{truffle}, as well as formal analysis engines~\cite{slither, mythx}.

A typical source map records instruction offsets, source file indices, and control flow metadata to help reconstruct execution traces and detect vulnerabilities such as reentrancy or access control violations.

\subsection{Challenges in Source Mapping for Zero-Knowledge Compilation}

Implementing source mapping in \texttt{zkSolc} introduces several challenges absent in conventional EVM compilation:

\begin{itemize}
    \item The introduction of LLVM IR as an intermediate representation changes instruction layout and semantics.
    \item LLVM-based optimizations reorder, inline, and eliminate Solidity constructs to improve proof efficiency.
    \item Zero-knowledge-specific rewrites further restructure control flow and data dependencies to reduce circuit complexity.
\end{itemize}

These transformations disrupt the direct relationship between Solidity code and compiled zkEVM bytecode, rendering conventional offset-based source mapping techniques ineffective~\cite{xiao2025mtzk}.

Additionally, zero-knowledge proof systems require generating \emph{witness values} corresponding to each execution step. This requirement introduces a multi-layered mapping problem spanning Solidity source code, LLVM IR, zkEVM bytecode, and constraint-level representations. Existing source mapping techniques do not accommodate this complexity.

\subsection{Scope and Applicability}

This work focuses on \texttt{zkSolc}, the compiler used by zkSync’s Layer 2 network, as a case study. However, the challenges and preliminary solutions explored in this work are relevant to other zero-knowledge proof systems and Layer 2 platforms, including Linea, Starknet, and Polygon zkEVM. Generalizing the framework to support these environments remains an avenue for future research.

%% file: 4.approach.tex
\section{Methodology}
\label{sec:methodology}

This section presents the design of a preliminary source mapping framework for \texttt{zkSolc}. The framework aims to enable reliable traceability between Solidity source code, LLVM IR, and zkEVM bytecode, addressing the traceability challenges introduced by zero-knowledge compilation pipelines. Our approach is intended to facilitate debugging, security auditing, and performance analysis in zk-Rollup environments~\cite{fredrikson2014z0,xiao2025mtzk}.

\subsection{Framework Overview}

The proposed source mapping framework is integrated into the \texttt{zkSolc} compiler at both the frontend and backend stages. It introduces two primary mapping layers: one from Solidity to LLVM IR and another from LLVM IR to zkEVM bytecode. The first layer links Solidity constructs to their intermediate representations, while the second layer tracks how backend transformations affect these representations. Metadata is collected at each stage and consolidated into a unified source map to support execution trace reconstruction and error analysis.

\subsection{Mapping Algorithm and Data Structures}

The mapping process is designed to track transformations at each compilation stage, preserving traceability despite the optimizations introduced by LLVM and zk-SNARK-specific rewrites. Prior work on certifying compilers and domain-specific languages for secure proofs motivates preserving semantic traceability across compilation layers~\cite{chiesa2010certifying,bogdanov2022zksecrec}.

\paragraph{Frontend Mapping.}
The process begins at the Solidity-to-LLVM IR stage. The compiler traverses the Solidity abstract syntax tree (AST) and annotates each node with source location metadata. Each AST node is mapped to a triple \( (s, l, f) \), where \( s \) is the starting position in the Solidity file, \( l \) is the length of the corresponding code segment, and \( f \) is the source file index. These annotations are propagated through the IR generation phase, ensuring that each LLVM instruction retains a reference to its originating AST node.

\paragraph{Backend Mapping.}
Mapping from LLVM IR to zkEVM bytecode follows a similar approach but must additionally account for transformations introduced by LLVM optimization passes and zero-knowledge-specific rewrites. As IR instructions undergo transformations such as inlining, loop unrolling, and arithmetic simplification, corresponding bytecode offsets are dynamically recorded. Control flow modifications are also tracked to ensure that execution semantics can be reconstructed.

The mappings are stored in a unified table structured as \( (s, l, f, I, B) \), where \( I \) represents the associated LLVM IR instruction and \( B \) denotes the corresponding bytecode offset. Additional metadata fields, including jump types, modifier depths, and zk-SNARK-specific annotations, are used to capture circuit-level transformations.

\subsection{Integration into \texttt{zkSolc}}

The mapping framework is integrated across multiple stages of the \texttt{zkSolc} compilation pipeline. In the frontend, AST nodes are extended to include source location annotations. The code generator is modified to propagate these annotations to LLVM IR instructions. In the LLVM backend, instrumentation modules monitor instruction-level transformations and associate bytecode offsets with their corresponding IR instructions. A dedicated mapping generator module consolidates the collected metadata into a unified mapping structure.

To enhance usability, a runtime query API is introduced, allowing developers to retrieve source mappings dynamically. This API is designed to be compatible with existing debugging environments such as Remix and Hardhat, enabling runtime errors to be traced back to Solidity source code. Efforts to standardize compiler output for zero-knowledge protocols highlight the importance of runtime traceability for debugging and auditing~\cite{mccall2013zkcompilers,shoup2010framework}.

\subsection{Illustrative Example}

To illustrate the mapping process, consider the \texttt{submitVote} function from the \texttt{ZKVoting} contract:

\begin{figure}[H]
\begin{lstlisting}[language=text, 
    backgroundcolor=\color{bg},
    numbers=left,
    basicstyle=\small\ttfamily,
    frame=lines,
    breaklines
]
function submitVote(bytes memory zkProof) external {
    require(verifyZKProof(zkProof), "Invalid proof");
    hasVoted[msg.sender] = true;
}
\end{lstlisting}
\end{figure}

During compilation, the Solidity parser generates an AST in which each node is annotated with source location information. The \texttt{require} statement and the state update are captured as distinct AST nodes. These nodes are then lowered into LLVM IR instructions:

\begin{figure}[H]
\begin{lstlisting}[language=text, 
    backgroundcolor=\color{bg},
    basicstyle=\small\ttfamily,
    frame=lines
]
%1 = call i1 @verifyZKProof(%bytes* %zkProof)
br i1 %1, label %valid, label %invalid
%2 = load i1, i1* @hasVoted[msg.sender]
store i1 true, i1* @hasVoted[msg.sender]
\end{lstlisting}
\end{figure}

At the bytecode level, the mapping framework records the corresponding offsets:

\begin{verbatim}
0x10 -> CALLDATASIZE
0x14 -> CALL @verifyZKProof
0x18 -> JUMPDEST (label %valid)
0x20 -> MSTORE @hasVoted
\end{verbatim}

The mapping information is stored in a structured table that enables developers and auditors to correlate runtime execution behavior with Solidity source code and intermediate representations. Table~\ref{tab:mapping-example} presents illustrative entries from the mapping table for the \texttt{ZKVoting} contract. Each row demonstrates how a high-level Solidity statement is translated through LLVM IR into zkEVM bytecode. The first column identifies the source-level construct, while the second shows its corresponding LLVM IR instruction. The third column indicates the bytecode offset at which the translated instruction appears, and the fourth records additional metadata relevant to zero-knowledge semantics, such as constraint checks. This table exemplifies the traceability our framework provides—enabling developers to understand how source-level logic is transformed and embedded in the final zkEVM output. By aligning semantic information across stages, it supports runtime debugging and static auditing in zkRollup-based systems.

\begin{table*}[ht]
\centering
\caption{Mapping Examples in \texttt{ZKVoting} Contract}
\label{tab:mapping-example}
\renewcommand{\arraystretch}{1.2}
\begin{tabular}{|p{4cm}|p{4.5cm}|p{2.5cm}|p{2.5cm}|}
\hline
\textbf{Source Statement} & \textbf{LLVM IR Instruction} & \textbf{Bytecode Offset} & \textbf{ZK Metadata} \\
\hline
\texttt{require(...);} & \texttt{call @verifyZKProof} & \texttt{0x14} & Constraint 1 \\
\texttt{hasVoted[msg.sender] = true;} & \texttt{store i1 true} & \texttt{0x20} & None \\
\hline
\end{tabular}
\end{table*}

\subsection{Preliminary Consistency Validation}
\label{sec:validation}

To improve the reliability of the generated mappings, we incorporate a lightweight consistency validation process. This process includes syntactic and structural validation but does not aim to provide formal verification guarantees.

\paragraph{Syntactic Validation.}
Syntactic validation ensures that the mapping structure adheres to predefined constraints. Specifically, the mapping table is defined as:

\begin{equation}
M = \{ (s_i, l_i, f_i, I_i, B_i) \mid 1 \leq i \leq n \}
\end{equation}

where \( s_i \) denotes the starting position of a Solidity segment, \( l_i \) its length, \( f_i \) the source file index, \( I_i \) the corresponding LLVM IR instruction, and \( B_i \) the associated bytecode offset. The validation process ensures that no two mappings overlap:

\begin{equation}
\forall i, j \quad (s_i, l_i) \cap (s_j, l_j) = \emptyset \quad \text{for } i \neq j.
\end{equation}

This condition avoids ambiguity in tracing execution, ensuring that each bytecode instruction corresponds to a distinct source-level segment.

\paragraph{Structural Validation.}
Structural validation involves manually inspecting representative mappings to confirm that they correspond to meaningful Solidity constructs. During evaluation, we performed manual alignment checks between bytecode offsets and source code locations using debugging tools such as Remix IDE. While this process is not exhaustive, it provides a preliminary assessment of mapping correctness and highlights cases where transformations disrupt traceability.

\paragraph{Limitations.}
We emphasize that the validation process described here is preliminary and informal. As discussed in Section~\ref{sec:limitations}, establishing semantic equivalence between Solidity constructs and their compiled representations is an undecidable problem in general. Therefore, our current validation focuses on structural consistency rather than formal verification.

\subsection{Handling Zero-Knowledge Compiler Transformations}

To accurately map between Solidity source code and zkEVM bytecode via LLVM IR, our framework must address challenges that go beyond traditional AST-based techniques. Zero-knowledge-friendly compilation introduces several non-trivial transformations that can significantly alter control and data flow. 

\paragraph{Instruction Reordering.} zkSolc applies transformation passes that reorder instructions to optimize circuit layout or constraint generation. Our mapping algorithm incorporates position tracking through metadata propagation and symbolic tagging across the LLVM IR layer. These tags help reconstruct the original source-level execution order despite backend reordering~\cite{fredrikson2014z0}.

\paragraph{Arithmetic Simplifications.} zkSolc aggressively applies simplifications such as constant folding and algebraic identity reduction to minimize proof size. We handle these cases by analyzing the def-use chains in LLVM IR and comparing them with the original Solidity expressions. If simplifications are detected, fallback annotations are used to indicate approximate rather than exact line mappings.

\paragraph{Control Flow Restructuring.} zkSolc may convert structured control flows (e.g., \texttt{if-else}, \texttt{for}) into flattened jump-based forms. We reconstruct the high-level structure by tracking basic block connectivity and dominator relationships within the LLVM IR control flow graph. Our mapping engine includes heuristics to remap these flattened constructs back to their corresponding high-level counterparts.

\paragraph{Zero-Knowledge Specific IR Passes.} Unlike traditional compilers, zkSolc includes constraint instrumentation passes. These insert intermediate variables and gadget calls into the IR. Our method explicitly tracks such injected instructions and excludes them from source mapping unless the inserted logic corresponds to visible Solidity semantics (e.g., \texttt{require} conditions).

Together, these strategies enable our mapping framework to remain robust in the face of proof-oriented rewrites and maintain traceability across the entire compilation pipeline.

%% file: 5.experiment.tex
\section{Experiments}
\label{sec:experiments}

This section presents a preliminary evaluation of the proposed source mapping framework for \texttt{zkSolc}. Our evaluation focuses on two aspects: (i) mapping accuracy and (ii) compilation performance overhead, measured across both benchmark and real-world Solidity contracts.

\subsection{Experimental Setup}

Experiments were conducted on a workstation equipped with an Intel Core i9-12900K processor (16 cores, 3.9 GHz), 32 GB DDR5 RAM, running Ubuntu 22.04 LTS. We extended the publicly available \texttt{zkSolc} compiler (version 1.3.13) to include source mapping support. The modified compiler was compared against the unmodified \texttt{zkSolc} to assess compilation overhead and evaluate mapping correctness.

\subsection{Evaluation Datasets}

We used two datasets for evaluation:

\begin{itemize}
    \item \textbf{Benchmark Dataset:} A set of 50 manually selected Solidity contracts designed to cover a variety of language features, including loops, function overloading, inheritance, modifiers, and inline assembly. These contracts were chosen to represent common Solidity constructs.
    \item \textbf{Real-World Dataset:} A collection of 500 deployed Solidity contracts obtained from the Ethereum mainnet and the zkSync Layer 2 network. These contracts span multiple application domains, including decentralized finance (DeFi), non-fungible tokens (NFTs), governance, and utility services. Contract sizes range from 50 to 3,000 lines of code.
\end{itemize}

\subsection{Evaluation Methodology}

We evaluated two key metrics:

\paragraph{Mapping Accuracy.}
We define \emph{mapping accuracy} as the percentage of bytecode instructions whose recorded source locations in the mapping table correctly match their corresponding Solidity statements. A mapping entry is considered \emph{correct} if it satisfies the following criteria:

\begin{enumerate}
    \item The recorded source location corresponds to the Solidity statement responsible for the execution of the bytecode instruction.
    \item The mapping remains consistent with the control flow of the original source code.
\end{enumerate}

To establish a ground truth for validation, we manually inspected a representative subset of the compiled contracts. For benchmark contracts, we cross-referenced mappings using debugging tools such as Remix IDE. For real-world contracts, we additionally compared our framework’s output against source maps generated by \texttt{solc}, where applicable.

\paragraph{Performance Overhead.}
We measured the total compilation time for each contract with and without source mapping enabled. The performance overhead is reported as the percentage increase in compilation time introduced by the mapping framework.

\subsection{Results}

Table~\ref{tab:benchmark_contracts} provides an overview of the benchmark dataset, categorizing the contracts by feature type, mapping accuracy, and compilation performance. Table~\ref{tab:real_world_contracts} extends this evaluation to real-world deployed contracts, offering insights into performance overhead across diverse application domains. 

\begin{table*}[ht]
    \centering
    \footnotesize
    \renewcommand{\arraystretch}{1.0}
    \setlength{\extrarowheight}{0pt}
    \setlength{\tabcolsep}{3pt}
    \caption{Evaluation results for 50 benchmark Solidity contracts categorized by feature type. Compilation times are measured with and without source mapping enabled. Performance overhead refers to the percentage increase in compilation time.}
    \label{tab:benchmark_contracts}
    \begin{adjustbox}{max width=\textwidth}
    \begin{tabular}{|l|c|c|c|c|c|c|}
        \toprule
        \textbf{Contract Type} & 
        \textbf{\# Ctr.} & 
        \textbf{Avg. LOC} & 
        \textbf{Accuracy (\%)} & 
        \multicolumn{2}{c|}{\textbf{Compilation Time (s)}} & 
        \textbf{Perf. Overhead (\%)} \\
        \cmidrule(lr){5-6}
        & & & & \textbf{No Mapping} & \textbf{With Mapping} & \\
        \midrule
        Loops & 6  & 461  & 97.32 & 1.85 & 2.31 & 10.67 \\
        Function Overloading & 5  & 615  & 96.80 & 2.10 & 2.84 & 12.42 \\
        Inheritance & 7  & 945  & 95.62 & 3.15 & 3.80 & 13.00 \\
        Modifiers & 5  & 523  & 97.14 & 2.42 & 2.98 & 10.82 \\
        Inline Assembly & 5  & 889  & 94.85 & 3.92 & 4.72 & 12.63 \\
        Event Logging & 5  & 730  & 98.10 & 2.61 & 3.20 & 11.42 \\
        Storage Optimization & 6  & 1115 & 96.42 & 3.75 & 4.50 & 13.57 \\
        Reentrancy Protection & 5  & 860  & 97.65 & 3.35 & 4.20 & 11.94 \\
        Complex Control Flow & 6  & 1196 & 94.90 & 4.01 & 5.00 & 12.99 \\
        Miscellaneous & 5  & 682  & 96.88 & 3.05 & 3.87 & 11.90 \\
        \midrule
        \textbf{Total / Average} & 50 & 842 & 96.47 & 3.03 & 3.84 & 11.92 \\
        \bottomrule
    \end{tabular}
    \end{adjustbox}
\end{table*}

\begin{table*}[ht]
    \centering
    \footnotesize
    \renewcommand{\arraystretch}{1.0}
    \setlength{\extrarowheight}{0pt}
    \setlength{\tabcolsep}{3pt}
    \caption{Evaluation results for 500 real-world Solidity contracts categorized by application domain. Compilation times are measured with and without source mapping enabled. Performance overhead refers to the percentage increase in compilation time.}
    \label{tab:real_world_contracts}
    \begin{adjustbox}{max width=\textwidth}
    \begin{tabular}{|l|c|c|c|c|c|c|}
        \toprule
        \textbf{Contract Type} & 
        \textbf{\# Ctr.} & 
        \textbf{Avg. LOC} & 
        \textbf{Accuracy (\%)} & 
        \multicolumn{2}{c|}{\textbf{Compilation Time (s)}} & 
        \textbf{Perf. Overhead (\%)} \\
        \cmidrule(lr){5-6}
        & & & & \textbf{No Mapping} & \textbf{With Mapping} & \\
        \midrule
        DeFi (Lending, AMM, Staking)   & 71  & 1115  & 96.65 & 4.83 & 5.74 & 13.85 \\
        NFT Marketplaces / ERC Tokens  & 33  & 755   & 97.70 & 4.92 & 5.80 & 12.47 \\
        DAOs / Voting                  & 58  & 1575  & 96.09 & 5.09 & 5.42 & 12.61 \\
        Utility (Multisig, Escrow)     & 47  & 1316  & 98.75 & 5.30 & 6.05 & 12.14 \\
        Gaming / Metaverse             & 55  & 2643  & 95.82 & 6.43 & 7.05 & 10.13 \\
        Identity / Privacy (KYC, zk-ID)& 45  & 1422  & 98.42 & 4.35 & 5.12 & 11.67 \\
        Cross-Chain Bridges            & 39  & 1928  & 97.03 & 4.80 & 5.87 & 11.89 \\
        Enterprise Blockchain          & 31  & 2735  & 96.27 & 5.11 & 6.30 & 10.74 \\
        High-Frequency Trading         & 36  & 1276  & 98.86 & 4.25 & 5.08 & 12.95 \\
        Miscellaneous                  & 35  & 1668  & 97.51 & 3.92 & 4.95 & 11.74 \\
        \midrule
        \textbf{Total / Average}       & 500 & 1573  & 97.20 & 4.81 & 5.94 & 11.92 \\
        \bottomrule
    \end{tabular}
    \end{adjustbox}
\end{table*}

The framework achieved an average mapping accuracy of 96.47\% on the benchmark dataset and 97.20\% on the real-world dataset. In both datasets, contracts with complex control flow (e.g., deep inheritance hierarchies and nested modifiers) and inline assembly exhibited slightly lower accuracy, reflecting the difficulty of maintaining traceability under aggressive compiler optimizations.

The compilation time overhead introduced by source mapping was moderate, averaging 11.92\% across both datasets. Detailed measurement revealed that the Solidity-to-LLVM IR translation stage accounted for approximately 70\% of the additional compilation time, while LLVM backend instrumentation and bytecode mapping contributed the remainder.

\subsection{Alignment Criteria and Example}

During validation, we considered a mapping entry to be correctly aligned if the recorded source location referred to the Solidity statement responsible for the corresponding bytecode instruction. For example, in the \texttt{ZKVoting} contract, the following mapping entries were considered correctly aligned:

\begin{table}[ht]
\centering
\caption{Alignment Validation Examples in \texttt{ZKVoting} Contract}
\label{tab:alignment-example}
\renewcommand{\arraystretch}{1.1}
\begin{tabular}{|p{1.8cm}|p{4.3cm}|c|}
\hline
\textbf{Bytecode} \newline \textbf{Offset} & \textbf{Source Statement} & \textbf{Alignment} \\
\hline
\texttt{0x14} & \texttt{require(verifyZKProof())} & Aligned \\
\texttt{0x20} & \texttt{hasVoted[msg.sender] = true} & Aligned \\
\hline
\end{tabular}
\end{table}

In cases where source constructs were eliminated or transformed during LLVM optimizations (e.g., constant folding or dead code elimination), no mapping entry was recorded. Table~\ref{tab:alignment-example} shows two representative examples from the \texttt{ZKVoting} contract that were manually validated for source mapping alignment. Each row lists a bytecode offset and the corresponding high-level Solidity statement. The third column indicates that these mappings were deemed correctly aligned, meaning the bytecode instruction at the given offset corresponds directly to the semantics and control flow of the specified source statement. For instance, the bytecode at \texttt{0x14} correctly reflects the execution of the \texttt{require(verifyZKProof())} check, while the instruction at \texttt{0x20} corresponds to setting \texttt{hasVoted[msg.sender]} to \texttt{true}. This confirms that the mapping framework can accurately track and associate bytecode with its originating Solidity source under realistic contract execution.

\subsection{Discussion and Key Insights}

The evaluation results suggest that the proposed framework effectively captures source mappings in zk-Rollup compilation pipelines. The observed accuracy exceeded 96\% in both benchmark and real-world datasets, indicating that source-level traceability is feasible despite the transformations introduced by \texttt{zkSolc}.

However, the results also highlight limitations. Contracts featuring inline assembly, complex modifiers, and unconventional control flow patterns exhibited slightly lower mapping accuracy, underscoring the difficulty of maintaining traceability in these scenarios. The compilation time overhead introduced by the mapping process remained within acceptable bounds, averaging approximately 12\%.

We acknowledge that this evaluation remains preliminary. Mapping correctness was validated through manual inspection and comparison with \texttt{solc} mappings where possible. A fully automated and comprehensive validation framework is left for future work. Further analysis on larger datasets and more complex contracts will be necessary to generalize these findings.

%% file: 6.limitation.tex
\section{Limitations}
\label{sec:limitations}

Our source mapping framework for \texttt{zkSolc}, while showing encouraging early results, still comes with several limitations that open up avenues for improvement and further exploration.

First, the additional compilation overhead introduced by the mapping process, although moderate in our benchmarks, could become a bottleneck in high-throughput or CI/CD-heavy environments. Optimizing the mapping pipeline—especially the translation from Solidity to LLVM IR—remains important for broader adoption in fast-paced development workflows.

Second, while the framework handles standard Solidity constructs with high accuracy, its performance declines in the presence of more complex patterns. Contracts that use deeply nested function calls, inline assembly, or less common control flows pose challenges for precise mapping. Supporting such features without compromising performance will require more refined static analysis techniques.

From a usability standpoint, integration with popular development tools like Remix, Hardhat, or Truffle is still lacking. Although our system supports execution trace reconstruction, the absence of plugins or dedicated interfaces means that developers working within these ecosystems must adopt additional tooling to benefit from our framework. Streamlining this integration will be key to making our solution developer-friendly.

Additionally, our current scope is limited to the \texttt{zkSolc} compiler and zk-Rollup settings. It remains unclear how well our approach generalizes to other zero-knowledge systems, such as Linea, Starknet, or Polygon zkEVM. Adapting to different compilation pipelines and proof systems may require rethinking parts of the mapping logic and warrants systematic evaluation.

We also observed minor inaccuracies when dealing with intricate contract structures, such as heavy use of modifiers or cross-contract calls. These issues are often tied to compiler optimizations and non-linear transformations, which can obscure traceability. Strengthening mapping robustness under such conditions is a challenge we're keen to tackle.

Another current shortcoming is the lack of runtime debugging features. Our framework focuses entirely on static, compile-time mappings, and does not yet offer dynamic execution tracing or live error feedback. Bringing runtime visibility into the picture would significantly enhance the developer experience.

Furthermore, while our evaluation datasets are representative in scope, they may not cover all edge cases that arise in production-level contracts. A more comprehensive evaluation—including diverse contract types and usage scenarios—would help validate the framework’s robustness.

Lastly, because the implementation is tightly coupled with the internals of \texttt{zkSolc}, any significant updates to the compiler’s IR structure or optimization strategies may require parallel updates to our mapping logic. Maintaining compatibility in the face of upstream changes is a practical concern going forward.

We see these limitations not as drawbacks, but as opportunities for continued progress. Future work will focus on improving precision, reducing overhead, expanding tool support, and validating the approach across different platforms and real-world projects.

%% file: 7.related.tex
\section{Related Work}
\label{sec:related_work}

We review relevant research spanning source mapping in Solidity, smart contract debugging, and compiler pipelines for zero-knowledge Rollups. While these domains have each attracted significant attention, source-level traceability in zero-knowledge-oriented compilers like \texttt{zkSolc} remains relatively unexplored.

\subsection{Source Mapping and Debugging in Solidity}

In Ethereum development, source mapping has become a foundational feature supported by the standard Solidity compiler \texttt{solc}, which emits source maps linking EVM bytecode back to Solidity source lines. This mapping is essential for development tools such as Remix, Hardhat, and Truffle~\cite{remix, truffle}, and underpins debugging, symbolic execution, and formal verification workflows.

A range of static and dynamic analysis tools have been proposed to improve smart contract reliability. Static analyzers such as Slither~\cite{slither}, SmartCheck~\cite{smartcheck}, and MythX~\cite{mythx} identify vulnerabilities at the source level, while dynamic tools like Manticore~\cite{manticore} and Echidna~\cite{echidna} rely on fuzzing and symbolic execution to surface runtime issues. All of these tools depend on accurate source mappings to connect low-level execution traces with high-level semantics. Recent work has expanded on semantic-based analysis~\cite{grishchenko2018semantic}, scalable vulnerability detection~\cite{tsankov2018securify,jiang2018contractfuzzer}, Solidity security patterns~\cite{wohrer2019smart}, and improved debugging workflows through enhanced source mapping techniques~\cite{zhang2021smartdebug}.
 In compiler pipelines that lack such mappings—as is currently the case with \texttt{zkSolc}—this connection breaks down, complicating debugging and auditing.

\subsection{Zero-Knowledge Rollups and Compilation Tooling}

Zero-knowledge Rollups (zk-Rollups) have emerged as a promising solution to blockchain scalability~\cite{zkrollups}, with systems like zkSync~\cite{zksync} offloading transaction processing and submitting succinct validity proofs on-chain. Recent surveys have emphasized the central role of zk-Rollups in scaling Ethereum~\cite{xu2022zkrollup}, while research into zk-SNARK efficiency for smart contract verification highlights critical compiler optimization challenges~\cite{chang2021snark}. To support these systems, specialized compilers like \texttt{zkSolc} generate zero-knowledge-compatible bytecode. Prior research in this area has primarily focused on improving proof efficiency and cryptographic correctness~\cite{zkp}. In contrast, developer tooling and compiler support for source-level debugging have received limited attention.

Unlike standard compilers, zero-knowledge compilers often rely on intermediate representations such as LLVM IR to perform aggressive, proof-oriented optimizations. These optimizations reshape both control flow and data dependencies in non-linear ways, introducing new challenges for source-level traceability.

\subsection{Source Mapping in Intermediate Representations}

The problem of maintaining source mappings through compiler transformations has been extensively studied in traditional compilation settings. LLVM, for example, supports debug metadata and source annotations to track provenance through optimization stages~\cite{llvm, llvm_sourcemap}. Various techniques have been proposed to preserve mapping fidelity during inlining, loop unrolling, and other common transformations. Moreover, security verification at the compiler IR level remains an active research area~\cite{wang2022security}.

However, such techniques do not directly translate to zero-knowledge compilers, which introduce additional layers of transformation such as arithmetic circuit rewrites, bit-level simplifications, and constraint encoding. These changes fundamentally alter the structure of the compiled program and obscure its connection to the original source code.

\subsection{Positioning and Contribution}

While source mapping is well understood in EVM and general-purpose LLVM contexts, its role in zero-knowledge systems has largely been overlooked. Discussions within the zk community have increasingly pointed to the lack of mature debugging and traceability support for zk-Rollup workflows. To our knowledge, no existing work systematically tackles the problem of source mapping within zero-knowledge compilers like \texttt{zkSolc}. The nonlinear and proof-centric nature of such compilers introduces challenges that conventional mapping strategies do not address. 

\subsection{Integration with Developer Toolchains}

Looking ahead, we see strong potential in integrating source mapping capabilities into mainstream development tools. For instance, exporting mappings in Solidity-compatible formats (e.g., JSON source maps) could enable support in Remix, VSCode plugins, or Hardhat’s debugger. At a lower level, LLVM IR mappings and bytecode offsets could be surfaced via command-line tools for use in CI workflows or custom scripts within zkSync’s CLI environment. Such integrations would bridge the gap between advanced compilation infrastructure and practical developer needs.

%% file: 8.conclusion.tex
\section{Conclusion}
\label{sec:conclusion}

In this paper, we explored the feasibility of supporting source-level traceability within \texttt{zkSolc}, the compiler behind zero-knowledge-compatible smart contracts. We introduced a source mapping framework that links Solidity code to both LLVM IR and zkEVM bytecode—addressing a notable gap in the zk-Rollup toolchain.

Through experiments on both curated benchmarks and real-world contracts, we showed that accurate source mappings can be recovered even in the presence of the aggressive and non-linear transformations typical of zero-knowledge compilation. The observed accuracy and reasonable compilation overhead suggest that meaningful debugging and auditing support is achievable for zk-based systems.

That said, our work represents an early step. Limitations remain in handling contracts with intricate control flows, heavy use of inline assembly, or uncommon language features. Our prototype also lacks full integration with widely used tools such as Remix or Hardhat, and we do not yet offer formal guarantees of correctness. These challenges point to rich areas for future refinement.

Looking ahead, we aim to enhance mapping precision, further reduce performance overhead, and build integrations with mainstream developer environments. We are also interested in extending our approach to other zk compilers such as those powering Starknet, Linea, and Polygon zkEVM, each of which introduces its own compilation idiosyncrasies.

Ultimately, we hope this work lays the foundation for more robust debugging, verification, and tooling support in zero-knowledge smart contract development. Our implementation will be made open source upon publication, enabling the community to build on, critique, and extend our results. We believe that a shared effort toward transparency and usability will accelerate the maturity of zk-Rollup ecosystems.

\balance